\documentclass[a4paper,11pt]{article}
\pdfoutput=1
\usepackage{amsmath,eurosym,amssymb,amsfonts,amsfonts,wasysym,latexsym, multirow, multicol,bm,color}
\usepackage{graphicx}
\RequirePackage[numbers,sort&compress]{natbib}
\def\bib{\bibitem}

\begin{document}

\renewcommand{\thefootnote}{\fnsymbol{footnote}}
\begin{center}
{\Large \bf A novel teleparallel dark energy model}
\vskip 0.5cm
{\bf  Giovanni Otalora}
\vskip 0.3cm
{\it Depto. de Matem\'atica, ICE, Universidade Federal de Juiz
de Fora, MG, Brazil  \\and\\ Instituto de F\'{\i}sica Te\'orica, UNESP-Universidade Estadual Paulista, \\
Caixa Postal 70532-2, 01156-970, SP, Brazil}
\vskip 0.8cm
\begin{quote}
{\bf Abstract.~}{\footnotesize Although equivalent to general relativity, teleparallel gravity is conceptually speaking a completely different theory. In this theory, the gravitational field is described by torsion, not by curvature. By working in this context, a new model is proposed in which the four-derivative of a canonical scalar field representing dark energy is nonminimally coupled to the ``vector torsion''. This type of coupling is motivated by the fact that a scalar field couples to torsion through its four-derivative, which is consistent with local spacetime kinematics regulated by the de Sitter group $SO(1,4)$. It is found that the current state of accelerated expansion of the Universe corresponds to a late-time attractor that can be (i) a dark-energy-dominated de Sitter solution ($\omega_{\phi}=-1$), (ii) a quintessence-type solution with $\omega_{\phi}\geq-1$, or (iii) a phantom-type $\omega_{\phi}<-1$ dark energy. 

}
\end{quote}
\end{center}
\vskip 0.8cm

\section{Introduction}
Like the other fundamental interactions of nature, gravitation can be described in terms of a gauge theory, the so-called \textit{Teleparallel Equivalent of General Relativity} or also known as Teleparallel Gravity (TG), which
attributes gravitation to torsion \cite{JGPereira, 12, SelectedtopicsinT, EMTensorTG}. A crucial concept of gravitation is that the metric tensor itself defines neither curvature nor torsion. 
In fact, curvature and torsion are properties of connections, and many different connections, with different curvature and torsion tensors, can be defined on the very same metric spacetime. 
A general Lorentz connection has $24$ independent components, and thus it is seen that any gravitational theory in which the source is the $10$ components symmetric energy-momentum tensor will not be able to determine uniquely the connection. 
The teleparallel connection and the Levi-Civita (or Christoffel) connection are the only two choices respecting the correct number of degrees of freedom of gravitation---all other choices will include additional degrees of freedom. 
The former may be considered a kind of  ``dual'' to the latter in the sense that, whereas the teleparallel connection has vanishing curvature and non-vanishing torsion, the Levi-Civita connection has vanishing torsion and non-vanishing curvature \cite{JGPereira,12,SelectedtopicsinT}.

On the other hand, from cosmic observations of Supernovae Ia (SNe Ia) \cite{obs1}, cosmic microwave background (CMB) radiation \cite{obs2}, large scale structure (LSS) \cite{obs3}, baryon acoustic oscillations (BAO) \cite{obs4}, and weak lensing \cite{obs5}, it is seen that the Universe is currently in a phase of accelerated expansion. Such phase is generally assumed to be driven by a peculiar form of energy, called dark energy, which in turn can be assumed to be generated by a scalar field with negative pressure. A cosmological constant is simpler and more natural than a scalar field, and could be considered as an alternative model. However, extreme fine tuning and coincidence problems make it quite problematic \cite{1,DarkETO}. In the context of modified gravity, other models have also been proposed, like for example $f(R)$ gravity \cite{f(R)gravity} and $f(T)$ gravity \cite{f(T)gravity}.

Considering that scalar fields can interact with other fields, such as the gravitational sector of the theory, and following the same spirit of scalar-tensor theories we can consider a nonminimal coupling between the scalar field and gravity. Many authors have studied models with a scalar field nonminimally coupled to gravity in the framework of GR \cite{Spokoiny,5,6,7,Barvinsky,NonMinimallyHiggs,8,9,Bertolami2000,Bertolami2010,NonMinimalTachyoninGR,16, 20,21,22,23,24,25,26}. Recently, it has been considered, in analogy with a similar construction in GR, a nonminimally coupled scalar field in the context of TG by adding a term $f(\phi)\,T$, with $f(\phi)$ a function of the scalar field and $T$ the so-called torsion scalar. This theory, which addresses the dark energy problem, has been called ``teleparallel dark energy'' (TDE) \cite{10,11,13,14, 39, 40,Tachyonic1, NonMinimalTachyon,41,42}. 

As is well-known, in the context of TG, although a scalar field itself does not feel gravity, its four-derivative (which is a vector field) interacts with the vector part of torsion \cite{JGPereira,12}. Inspired in this property, the purpose of this paper is to study a new dark energy model in which the four-derivative of the scalar field couples nonminimally to the ``vector torsion''. In Ref. \cite{HJennenJGPereira}, it has been shown that this class of nonminimal coupling naturally emerges in the context a generalized teleparallel gravity, ``de Sitter teleparallel gravity'',  which is consistent with local spacetime kinematics regulated by the de Sitter group $SO(1,4)$. Throughout the paper we adopt natural units $c=1$ such that $\kappa^{2}=8 \pi G$; we use a metric with signature $(+,-,-,-)$.

\section{The model}
The torsion tensor can be decomposed into three components, irreducible under the global Lorentz group:
there will be a vector 
\begin{equation}
\mathcal{V}_{\mu}=T^{\nu}_{~\nu \mu},
\label{5.3}
\end{equation}an axial part
\begin{equation}
 \mathcal{A}^{\mu}=\frac{1}{6}\,\epsilon^{\mu \nu \rho \sigma}\,T_{\nu \rho \sigma},
 \label{5.4}
\end{equation}and a purely tensor part
\begin{equation}
 \mathcal{T}_{\lambda \mu \nu}=\frac{1}{2}\,\left(T_{\lambda \mu \nu}+T_{\mu \lambda \nu}\right)+\frac{1}{6}\,\left(g_{\nu \lambda}\,\mathcal{V}_{\mu}+g_{\nu \mu}\,\mathcal{V}_{\lambda}\right)-\frac{1}{3}\,g_{\lambda \mu}\,\mathcal{V}_{\nu},
 \label{5.5}
\end{equation}that is, a tensor with vanishing vector and axial parts. These components are usually called ``vector torsion'', ``axial torsion'' and ``pure tensor torsion'' \cite{JGPereira}. Since a scalar field interacts with torsion through its four-derivative, then the four-divergence of the scalar field can be nonminimally coupled with the vector torsion.
So, let us consider the following action for the nonminimally coupled quintessence field 
\begin{equation}
 S=\int d^{4}x\,h\,\left[\frac{T}{2\,\kappa^{2}}+\frac{1}{2}\partial_{\mu}{\phi}\,\partial^{\mu}{\phi}-V(\phi)+\eta\,f(\phi)\,\partial_{\mu}{\phi}\,\mathcal{V}^{\mu}\right]+S_{m}(\psi_{m},h^{a}_{~\rho}),
 \label{5.6}
\end{equation} where $h\equiv\det(h^{a}_{~\mu})=\sqrt{-g}$, $\mathcal{L}_{G}=\frac{h\,T}{2\,\kappa^2}$ is the lagrangian of TG and $S_{m}(\psi_{m},h^{a}_{~\rho})$ is the matter action \cite{JGPereira,12,14}. The parameter $\eta$ is a dimensionless constant and $f(\phi)$ is a function of the scalar field with units of $mass$. 

The energy momentum tensor associated with the scalar field is calculated as
\begin{multline}
 \Theta_{a}^{~\rho}\equiv-\frac{1}{h}\frac{\delta{S}_{\phi}}{\delta{h^{a}_{~\rho}}}=\eta\,\left[f(\phi)\,\left(\mathcal{V}^{\rho}\,\partial_{a}{\phi}+\nabla_{a}{\partial^{\rho}{\phi}}-h_{a}^{~\rho}\,\nabla_{\mu}{\partial^{\mu}{\phi}}\right)+f_{,\phi}\,\left(\partial_{a}{\phi}\,\partial^{\rho}{\phi}-h_{a}^{~\rho}\,\partial_{\mu}{\phi}\,\partial^{\mu}{\phi}\right)\right]-\\
 h_{a}^{~\rho}\,\left(\frac{1}{2}\,\partial_{\mu}{\phi}\,\partial^{\mu}{\phi}-V(\phi)\right)+\partial_{a}{\phi}\,\partial^{\rho}{\phi}, 
 \label{5.11}
\end{multline} where $\nabla^{\mu}$ is the covariant derivative in the teleparallel connection \cite{JGPereira,12} and $f_{,\phi}\equiv\frac{df}{d\phi}$. 
The symmetric part is given by
\begin{multline}
\Theta_{(\mu \nu)}=\eta\,[f(\phi)\,\left(\mathcal{V}_{(\mu}\,\partial_{\nu)}{\phi}+\nabla_{(\mu}{\partial_{\nu)}{\phi}}-g_{\mu\nu}\,\nabla_{\epsilon}{\partial^{\epsilon}{\phi}}\right)+f_{,\phi}\,\left(\partial_{\mu}{\phi}\,\partial_{\nu}{\phi}-g_{\mu \nu}\,\partial_{\epsilon}{\phi}\,\partial^{\epsilon}{\phi}\right)]-\\
g_{\mu \nu}\,\left(\frac{1}{2}\,\partial_{\epsilon}{\phi}\,\partial^{\epsilon}{\phi}-V(\phi)\right)+\partial_{\mu}{\phi}\,\partial_{\nu}{\phi},
 \label{5.12}
\end{multline} whereas the anti-symmetric part is 
\begin{equation}
\Theta_{[\mu \nu]}=\frac{2\,\kappa^{2}}{h}\,\eta\,f(\phi)\,\partial_{\epsilon}{\phi}\,\sigma^{\epsilon}_{~\mu\nu}, 
 \label{5.13}
\end{equation}where $\sigma^{\rho}_{~\mu \nu}$ is the {\it Spin Tensor} of the gravitational field, which is defined as 
\begin{equation}
 \sigma^{\mu}_{~\lambda \gamma}\equiv-\frac{\partial{\mathcal{L}_{G}}}{\partial{\partial_{\mu}{h^{a}_{~\sigma}}}}\,\frac{\delta{h^{a}_{~\sigma}}}{\delta{\epsilon^{\lambda\gamma}}}=\frac{h}{\kappa^{2}}\,S_{[\lambda\gamma]}^{~~~~\mu}=-\frac{h}{4\,\kappa^{2}}\left(T^{\mu}_{~\lambda \gamma}+\delta^{\mu}_{~\gamma}\,\mathcal{V}_{\lambda}-\delta^{\mu}_{~\lambda}\,\mathcal{V}_{\gamma}\right),
 \label{C.16}
\end{equation}
with $\delta{\epsilon^{\alpha \beta}}$ an infinitesimal anti-symmetric (Lorentz) tensor and $S_{[\lambda \gamma]}^{~~~~\mu}$ the anti-symmetric part of the superpotential \cite{ClassicalFields, JGPereira}. However, the antisymmetric part \eqref{5.13} is not relevant on cosmological scales where there is homogeneity and isotropy.
Varying the action with respect to the scalar field we find the motion equation
\begin{equation}
\nabla_{\mu}{\partial^{\mu}{\phi}}-\partial_{\mu}{\phi}\,\mathcal{V}^{\mu}+\eta\,f(\phi)\,\left(\nabla_{\mu}{\mathcal{V}^{\mu}}-\mathcal{V}_{\mu}\,\mathcal{V}^{\mu}\right)+V_{,\phi}=0.
\label{5.14}
\end{equation}which is written in terms of the covariant derivative of the teleparallel connection and $V_{,\phi}\equiv\frac{dV}{d\phi}$.

By imposing the flat FLRW geometry  
\begin{equation}
 h^{a}_{~\mu}(t)= \mbox{diag}(1,a(t),a(t),a(t)),
 \label{1.15}
\end{equation} we obtain for the energy density
\begin{equation}
 \rho_{\phi}=\frac{1}{2}\,\dot{\phi}^{2}+V(\phi)-3\,\eta\,f(\phi)\,H\,\dot{\phi},
 \label{5.15}
\end{equation} and for the pressure density
\begin{equation}
 p_{\phi}=\frac{1}{2}\,\left(1+2\,\eta\,f_{,\phi}\right)\,\dot{\phi}^{2}-V(\phi)+\eta\,f(\phi)\,\ddot{\phi}.
 \label{5.16}
\end{equation}On the other hand, imposing the same background \eqref{1.15} in the motion equation \eqref{5.14} we find
\begin{equation}
 \ddot{\phi}+3\,H\,\dot{\phi}-3\,\eta\,\left(\dot{H}+3\,H^{2}\right)\,f(\phi)+V_{,\phi}=0.
 \label{5.17}
\end{equation} This is the evolution equation for the scalar field and can alternatively be written in the standard form 
$\dot{\rho}_{\phi}+3\,H\,\left(1+\omega_{\phi}\right)\,\rho_{\phi}=0$ with $\omega_{\phi}\equiv\frac{p_{\phi}}{\rho_{\phi}}$ the equation-of-state parameter.
\section{Cosmological dynamics}
To study the cosmological dynamics of the model, we introduce the followings dimensionless variables:
\begin{equation}
x\equiv\frac{\kappa\,\dot{\phi}}{\sqrt{6}\,H},\:\:\:\:\:\:\:\ y\equiv\frac{\kappa\,\sqrt{V}}{\sqrt{3}\,H}, \:\:\:\:\:\:\:\:\: u\equiv\kappa\,f,\:\:\:\:\:\:\:\:\:\:\: \lambda\equiv-\frac{V_{,\phi}}{\kappa\,V},\:\:\:\:\:\:\:\:\:\: \alpha\equiv f_{,\phi}.
\label{5.18}
\end{equation} In terms of these dimensionless variables, the fractional energy densities $\Omega_{\phi}$ and $\Omega_{m}$ for the scalar field and background matter are given by
\begin{equation}
 \Omega_{\phi}\equiv\frac{\kappa^{2}\rho_{\phi}}{3H^{2}}={x}^{2}+{y}^{2}-\sqrt{6}\,\eta\,u\,x, \:\:\:\:\:\:\:\: \Omega_{m}\equiv\frac{\kappa^{2}\rho_{m}}{3H^{2}}=1-\Omega_{\phi},
 \label{5.19}
\end{equation}respectively. By using the physical condition $0\leq\Omega_{\phi}\leq1$ in equation \eqref{5.19}, the range in the phase space for the variables $x$, $u$ and $y$ is constrained. 

On the other hand, the equation of state of the field $\omega_{\phi}$ reads
\begin{equation}
 \omega_{\phi}=\frac{\left(1+2\,\eta\,\alpha\right)\,x^{2}-y^{2}+\eta\,u\,\left(-\sqrt{6}\,x+\eta\,u\,\left(3-s\right)+\lambda\,y^{2}\right)}{{x}^{2}+{y}^{2}-\sqrt{6}\,\eta\,u\,x}.
 \label{5.20}
\end{equation}
The effective equation of state $\omega_{eff}$ is given by
\begin{multline}
\omega_{eff}\equiv\frac{p_{m}+p_{\phi}}{\rho_{m}+\rho_{\phi}}=\left(\gamma-1\right)\,\left[1-\left(x^{2}+y^{2}-\sqrt{6}\,\eta\,u\,x\right)\right]+\left(1+2\,\eta\,\alpha\right)\,x^{2}-y^{2}+\\
\eta\,u\,\left(-\sqrt{6}\,x+\eta\,u\,\left(3-s\right)+\lambda\,y^{2}\right),
\label{5.21}
\end{multline}where we have defined $\gamma\equiv1+\omega_{m}$ and the accelerated expansion occurs for $\omega_{eff}<-\frac{1}{3}$. Also, it is defined the parameter
\begin{multline}
s\equiv-\frac{\dot{H}}{H^{2}}=\left(\frac{2}{3}+\eta^{2}\,u^{2}\right)^{-1}\,[\gamma-\left(\gamma-1\right)\,\left(x^{2}+y^{2}-\sqrt{6}\,\eta\,u\,x\right)+\left(1+2\,\eta\,\alpha\right)\,x^{2}+\\
\left(\eta\,\lambda\,u-1\right)\,y^{2}+\sqrt{6}\,\eta\,u\,\left(\frac{\sqrt{6}}{2}\,\eta\,u-x\right)].
\label{5.22}
\end{multline}
The dynamical system of ordinary differential equations (ODE) for the model is written as 
\begin{equation}
 x'=\left(3-s\right)\,\left(-x+\frac{\sqrt{6}}{2}\,\eta\,u\right)+\frac{\sqrt{6}}{2}\,\lambda\,y^{2},
 \label{5.23}
\end{equation}
\begin{equation}
 y'=\left(s-\frac{\sqrt{6}\,\lambda\,x}{2}\right) \,y,
 \label{5.24}
\end{equation}
\begin{equation}
 u'=\sqrt{6}\,\alpha\,x,
 \label{5.25}
\end{equation}
\begin{equation}
 \lambda'=-\sqrt{6}\,\left( \Gamma-1\right) \,{\lambda}^{2}\,x,
 \label{5.26}
\end{equation}
\begin{equation}
 \alpha'=\sqrt{6}\,\Pi\,x.
 \label{5.27}
\end{equation}
In these equations primes denote derivative with respect to the so-called e-folding time $N\equiv\ln a$ \cite{1,DarkETO}. Also, we have defined the parameters
\begin{equation}
 \Pi\equiv\kappa^{-1}\,f_{,\phi\phi},\:\:\:\:\:\:\:\:\:\: \Gamma\equiv\frac{V\,V_{,\phi\phi}}{V_{,\phi}^{2}}.
 \label{5.28}
\end{equation}
From now we concentrate on exponential scalar field potential of the form $V(\phi)=V_{0}\,e^{-\lambda\,\kappa\,\phi}$, such that $\lambda$ is a dimensionless constant, that is, $\Gamma=1$ (equivalently, we could consider potentials satisfying $\lambda\equiv-\frac{V_{,\phi}}{\kappa\,V} \thickapprox const$, which is valid for arbitrary but nearly flat potentials \cite{27,28,29}).
We study two simple cases: one where $\alpha$ is a constant and  another where $\alpha$ depends on $u$. In the first case, it is considered a nonminimal coupling function $f(\phi)\propto\phi$ and thus $\alpha=const$ and $\Pi=0$. Finally, we consider a dynamically changing $\alpha(u)$. If $u(\phi)\equiv\kappa\,f(\phi)$ is a general function, with inverse function $\phi(u)=f^{-1}(u/\kappa)$ thus $\alpha(\phi)$ and $\Pi(\phi)$  can be expressed in terms of $u$ (see \cite{30, 14, NonMinimalTachyon}). 
In this form, the dynamical system of (ODE) \eqref{5.23}-\eqref{5.27} is a dynamical autonomous system and we can obtain the fixed points or critical points $(x_{c},y_{c},u_{c})$  by imposing the conditions $x'_{c}=y'_{c}=u'_{c}=0$. 
From the definition \eqref{5.18}, $x_{c}$, $y_{c}$, $u_{c}$  should be real, with $y_{c}\geq0$. To study the stability of the critical points, we substitute linear perturbations, $x\rightarrow x_{c}+\delta{x}$, $y\rightarrow y_{c}+\delta{y}$, and $u\rightarrow u_{c}+\delta{u}$ around each critical point and linearize them. 
The eigenvalues of the perturbations matrix $\mathcal{M}$, namely, $\mu_{1}$, $\mu_{2}$ and $\mu_{3}$,  determine the conditions of stability of the critical points \cite{1}: (i) Stable node: $\mu_{1}<0$, $\mu_{2}<0$ and $\mu_{3}<0$. (ii) Unstable node: $\mu_{1}>0$, $\mu_{2}>0$ and $\mu_{3}>0$. (iii) Saddle point: one or two of the three eigenvalues are positive and the other negative. (iv) Stable spiral: The determinant of the matrix
$\mathcal{M}$ is negative and the real parts of $\mu_{1}$, $\mu_{2}$ and $\mu_{3}$ are negative.  A  critical point is an attractor in the cases (i) and (iv), but it is not so in the cases (ii) and (iii). The Universe will eventually enter these
attractor solutions regardless of the initial conditions.

\begin{table}[ht!]
 \centering
 \caption{Critical points for $\alpha=const$ and stability properties.}
\begin{center}
\begin{tabular}{c c c c c c c c c c }\hline \hline
Name & $x_{c}$ & $y_{c}$ & $u_{c}$  & $\Omega_{\phi}$ & $\omega_{\phi}$ &$\omega_{eff}$ & Existence & Acceleration & Stability\\\hline
I.a & $0$ & $0$ & $0$ &  $0$ & $-1$ & $\gamma-1$   & All values  & No&  SP or UN \\\hline
I.b & $0$ & $0$ & $u_{c}$ &  $0$ & $\eta\,\lambda\,u_{c}-1$ & $1$   & $\omega_{m}=1$  & No  & Unstable\\\hline
I.c & $0$ & $1$ & $-\frac{\lambda}{3\,\eta}$ &  $1$ & $-1$ & $-1$  &  All values & All values & SN or SS or SP \\\hline \hline
\end{tabular}
\end{center}
\label{table1}
\end{table}
\section{Constant $\alpha$}
\subsection{Critical points}
In this section we consider a nonminimal coupling function $f(\phi)\propto\phi$ such that $\alpha$ is a constant. The critical points are presented in the Table \ref{table1}. The critical point I.a is a matter-dominated solution ($\Omega_{m}=1$) with equation of state type cosmological constant $\omega_{\phi}=-1$, that exists for all values. 
The point I.b is also  a matter-dominated solution that exists for $u_{c}\in\Re$  and $\omega_{m}=\omega_{eff}=1$. On the other hand, the fixed point I.c correspond to a dark-energy-dominated de Sitter solution  with $\Omega_{\phi}=1$ and $\omega_{\phi}=\omega_{eff}=-1$. This point exists for all values.
It is a viable cosmological solution to describe the current accelerated expansion of the Universe.

\begin{figure}[ht!]
\centering
\includegraphics[width=0.45\textwidth]{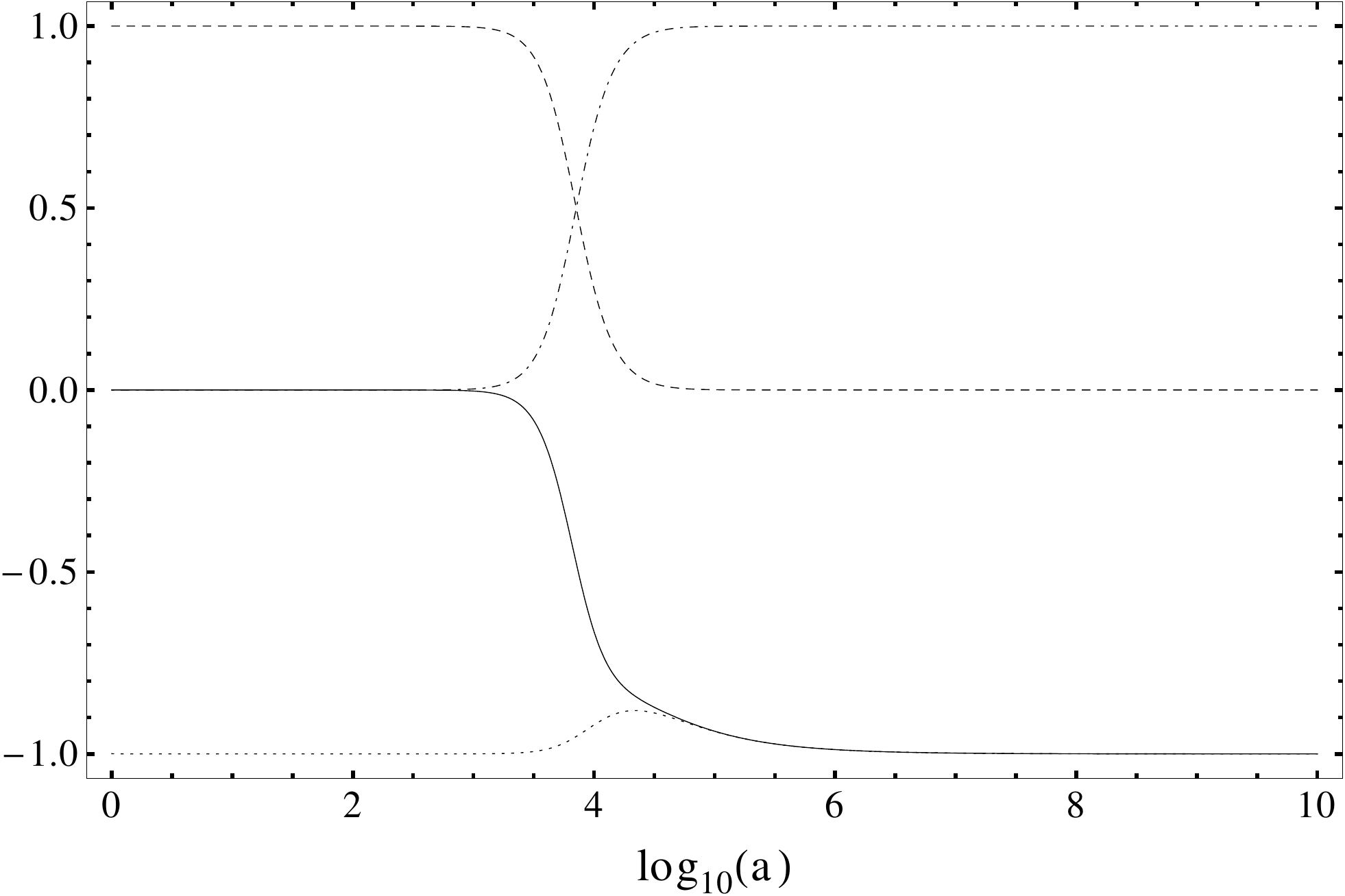}
\caption{Evolution of $\Omega_{m}$ (dashed), $\Omega_{\phi}$ (dotdashed), $\omega_{\phi}$ (dotted) and  $\omega_{eff}$ (solid) with $\gamma=1$,  $\lambda=0.8$, $\alpha=1$ and $\eta=-0.2$. The system asymptotically evolves toward the values $\Omega_{\phi}=1$, $\Omega_{m}=0$ and $\omega_{\phi}=\omega_{eff}=-1$.
Also, we have $\Omega_{\phi}\approx0.72$, $\Omega_{m}\approx0.28$, $\omega_{\phi}\approx-0.92$ and $\omega_{eff}=-0.66$ at the present epoch $N=\log_{10}{(a)}\approx4$.}
\label{figure1}
\end{figure}
\subsection{Stability}
Substituting the  linear perturbations, $x\rightarrow x_{c}+\delta{x}$, $y\rightarrow y_{c}+\delta{y}$, and $u\rightarrow u_{c}+\delta{u}$  into the autonomous system \eqref{5.23}-\eqref{5.27} and linearize them, we calculate the matrix of linear perturbations $\mathcal{M}$. The components of $\mathcal{M}$ are shown in the Appendix A.1.
The eigenvalues of $\mathcal{M}$ for each critical point are given by
\begin{itemize}
 \item Point I.a:
\end{itemize}
\begin{equation}
 \mu_{1,2}=\frac{3}{4}\,\left(2-\gamma\right)\,\left(-1\pm\sqrt{1+\frac{8\,\eta\,\alpha}{2-\gamma}}\right),\:\:\:\:\:\:\:\: \mu_{3}=\frac{3\,\gamma}{2}.
 \label{5.37}
\end{equation}
\begin{itemize}
 \item Point I.b:
\end{itemize}
\begin{equation}
\mu_{1,2}=0,\:\:\:\:\:\:\:\: \mu_{3}=3.
\label{5.38}
\end{equation}
\begin{itemize}
 \item Point I.c:
\end{itemize}
\begin{equation}
\mu_{1,2}=\frac{3}{2}\,\left(-1\pm\sqrt{1+\frac{24\,\eta\,\alpha}{\lambda^{2}+6}}\right),\:\:\:\:\:\:\:\: \mu_{3}=-3\,\gamma.
\label{5.39}
\end{equation}
The point I.a is always unstable, either saddle point (SP) or unstable node (UN). The point I.b. is also unstable for all values. 
Finally, the critical point I.c is a stable node (SN) for
\begin{equation}
 -\frac{\lambda^{2}+6}{24}\leq\eta\,\alpha<0.
 \label{5.40}
\end{equation}For $\eta\,\alpha>0$ it is a saddle point (SP). On the other hand, when 
\begin{equation}
 \eta\,\alpha<-\frac{\lambda^{2}+6}{24},
 \label{5.41}
\end{equation}thus $\mu_{1}$ and $\mu_{2}$ are complex with real part negative. 
In this case, the determinant of the matrix of perturbations $\det{\mathcal{M}}|_{(x_{c},y_{c},u_{c})}=\frac{162\,\gamma\,\eta\,\alpha}{\lambda^{2}+6}$ is negative and the point I.c is a stable spiral (SS).  Thus, we find that the fixed point I.c is an attractor (for conditions \eqref{5.40} or \eqref{5.41}) and a viable cosmological solution to explain the late-time accelerated expansion. The Universe will eventually enter this solution regardless of the initial conditions. In the Fig. \ref{figure1} it is shown as the Universe tends asymptotically to the dark-energy-dominated de Sitter solution I.c, passing through the matter-dominated solution I.a. 

\begin{table}[ht!]
 \centering
 \caption{Critical points for dynamically changing $\alpha(u)$. }
\begin{center}
\begin{tabular}{c c c c c c c}\hline \hline
Name & $x_{c}$ & $y_{c}$ & $u_{c}$ & $\Omega_{\phi}$ & $\omega_{\phi}$ &$\omega_{eff}$\\\hline
II.a & $\frac{\sqrt{6}\,\eta}{2}-\sqrt{1+\frac{3\,\eta^{2}}{2}}$ & $0$ &  $1$ &  $1$ & $1$ & $1$\\\hline
II.b & $\frac{\sqrt{6}\,\eta}{2}+\sqrt{1+\frac{3\,\eta^{2}}{2}}$ & $0$  & $1$ &  $1$ & $1$ & $1$\\\hline
II.c & $\frac{\sqrt{6}\,\gamma}{2\,\lambda}$ & $\frac{\sqrt{3}\,\sqrt{2-\gamma}\,\sqrt{\gamma-\eta\,\lambda}}{\sqrt{2}\,\left| \lambda\right| }$  & $1$ &  $\frac{3\,\left( \gamma\,\left( 2-\eta\,\lambda\right)-2\,\eta\,\lambda\right) }{2\,{\lambda}^{2}}$ & $\gamma-1$ & $\gamma-1$\\\hline
II.d & $\frac{\sqrt{6}\,\left(3\,\eta+\lambda\right) }{3\,\left(2-\eta\,\lambda\right) }$ & $\frac{\sqrt{6\,\left( 1-\eta\,\lambda\right)-{\lambda}^{2}}\,\sqrt{\frac{2}{3}+{\eta}^{2}}}{2-\eta\,\lambda}$  & $1$ & $1$ & \eqref{5.42} &  \eqref{5.42} \\\hline \hline
\end{tabular}
\end{center}
\label{table3}
\end{table}
\begin{table}[ht!]
\caption{Stability properties, and conditions for acceleration and existence of the fixed points in Table \ref{table3}.}
 \centering
\begin{center}
\begin{tabular}{c c c c c c c}\hline \hline
Name & Stability & Acceleration & Existence \\\hline 
II.a &  UN or SP ($\gamma=1$)  &  No  & All values \\\hline
II.b &  UN or SP ($\gamma=1$) &  No  & All values \\\hline
II.c &  SN or SP or SP ($\gamma=1$) &  No  & $\frac{2\,\left(3\,\gamma-\lambda^2\right)}{3\,\left(\gamma+2\right)}\leq\eta\,\lambda\leq\frac{2\,\gamma}{\gamma+2}$\\\hline
II.d &  SN or SP  &  $\eta\,\lambda<\frac{1}{2}-\frac{\lambda^{2}}{4}$  & $\eta\,\lambda\leq 1-\frac{\lambda^{2}}{6}$ \\\hline \hline
\end{tabular}
\end{center}
\label{table4}
\end{table}

\section{Dynamically changing $\alpha$}
\subsection{Critical points}
Following Refs. \cite{30, 14, NonMinimalTachyon}, let us consider a general nonminimal coupling function $f(\phi)$ such that $\alpha$ can be expressed in terms of $u$ and $\alpha(u)\rightarrow\alpha(u_{c})=0$  when $(x,y,u)\rightarrow (x_{c},y_{c},u_{c})$.
The field $\phi$ rolls down toward $\pm \infty$ ($x>0$ or $x<0$) with $f(\phi)\rightarrow 1/\kappa$ and $u_{c}=1$. The fixed points are presented in the Table \ref{table3}, and the properties in Table \ref{table4}.

The fixed points II.a and II.b are both scalar-field-dominated solutions with $\Omega_{\phi}=1$ and equation of state type ``stiff matter'' $\omega_{\phi}=1$. 
These points exist for all values. The fixed point II.c is a scaling solution that exists for $\eta\,\lambda<\gamma<2$. This point is a realistic solution when
\begin{equation}
\frac{2\,\left(3\,\gamma-\lambda^2\right)}{3\,\left(\gamma+2\right)}\leq\eta\,\lambda\leq\frac{2\,\gamma}{\gamma+2},
\label{PhysicalConditionPointII.d}
\end{equation} since in this case $0\leq\Omega_{\phi}\leq1$. For nonrelativistic matter $\gamma=1$ thus $\frac{2}{3}-\frac{2\,\lambda^{2}}{9}\leq\eta\,\lambda\leq\frac{2}{3}$. On the other hand, in the case of relativistic matter (radiation) $\gamma=4/3$ we have that $\frac{4}{5}-\frac{\lambda^{2}}{5}\leq\eta\,\lambda\leq\frac{4}{5}$. 
Just like fixed points II.a, II.b, the  point II.c is not viable to explain a late-time acceleration. However, since point II.c is a scaling solution, this can be used to provide the cosmological evolution in which the energy density of the scalar field decreases proportionally to that of the background fluid in either a radiation or matter-dominated era.
Finally, the point II.d is also a scalar-field-dominated solution, but different to II.a and II.b, this point is a viable solution to explain the late-time cosmic acceleration. This point exists for $\eta\,\lambda\leq 1-\frac{\lambda^{2}}{6}$. The equation of state is given by
\begin{equation}
 \omega_{\phi}=\omega_{eff}=\frac{2\,{\lambda}^{2}+3\,(3\,\eta\,\lambda-2)}{3\,\left(2- \eta\,\lambda\right)},
 \label{5.42}
\end{equation}and the accelerated expansion occurs for $\eta\,\lambda<\frac{1}{2}-\frac{\lambda^{2}}{4}$. From \eqref{5.42} we have that $\omega_{\phi}\geq-1$ if $\eta\,\lambda\geq-\frac{\lambda^{2}}{3}$. Moreover, it is also possible that this solution to be phantom, that is $\omega_{\phi}<-1$, if $\eta\,\lambda<-\frac{\lambda^{2}}{3}$.
Is worth highlighting that unlike the dark energy models with phantom or ghost scalar field \cite{1, DarkETO}, the present model is devoid of any quantum instability \cite{QuantumInstability}.

\subsection{Stability}
The components of the matrix of linear perturbations $\mathcal{M}$ are shown in the Appendix A.2.
The eigenvalues of $\mathcal{M}$ for each critical point are as follows
\begin{itemize}
 \item Point II.a ($\gamma=1$):
\end{itemize}
\begin{equation}
\mu_{1}=\frac{\lambda}{2}\,\left(\sqrt{9\,\eta^{2}+6}-\left(3\,\eta-\frac{6}{\lambda}\right)\right),\:\:\:\:\:\:\:\: \mu_{2}=3, \:\:\:\:\:\:\:\:\:\: \mu_{3}=\tau_{c}\,\left(3\,\eta-\sqrt{9\,\eta^{2}+6}\right).
\label{5.51}
\end{equation}
\begin{itemize}
 \item Point II.b ($\gamma=1$):
\end{itemize}
\begin{equation}
\mu_{1}=-\frac{\lambda}{2}\,\left(\sqrt{9\,\eta^{2}+6}+3\,\eta-\frac{6}{\lambda}\right),\:\:\:\:\:\:\:\: \mu_{2}=3,\:\:\:\:\:\:\:\:\:\: \mu_{3}=\tau_{c}\,\left(3\,\eta+\sqrt{9\,\eta^{2}+6}\right).
\label{5.52}
\end{equation}
\begin{itemize}
 \item Point II.c ($\gamma=1$):
\end{itemize}
\begin{multline}
\mu_{1,2}=\frac{3}{4}\,\left(-1\pm\sqrt{1+Y}\right),\:\:\:\:\:\:\: \mu_{3}=\frac{3\,\tau_{c}}{\lambda},\:\:\:\:\:\:\:\:\:\: Y=\frac{8\,\left(\eta\,\lambda-1\right)\,\left(2\,\lambda^{2}+9\,\eta\,\lambda-6\right)}{\left(3\,\eta^{2}+2\right)\,\lambda^{2}}.
\label{5.53}
\end{multline}
\begin{itemize}
 \item Point II.d:
\end{itemize}
\begin{equation}
\mu_{1}=\frac{3\,\left(\gamma+2\right)\,\eta\,\lambda+2\,\left(\lambda^{2}-3\,\gamma\right)}{2-\eta\,\lambda},\:\:\:\:\:\:\:\: \mu_{2}=\frac{6\,\eta\,\lambda+\lambda^{2}-6}{2-\eta\,\lambda}, \:\:\:\:\:\:\: \mu_{3}=\frac{2\,\left(\lambda+3\,\eta\right)\,\tau_{c}}{2-\eta\,\lambda}.
\label{5.54}
\end{equation} 

Here $\tau_{c}$ is defined by $\tau_{c}\equiv\frac{d\alpha(u)}{du}|_{u=u_{c}}$. When the background fluid is nonrelativistic matter $\gamma=1$, the critical points II.a and II.b are unstable in any case. The scaling solution II.c, for $\gamma=1$ and $\eta\,\lambda<2/3$, is a stable node (SN) if
\begin{equation}
 \frac{3\,\left(2-3\,\eta\,\lambda\right)}{2}<\lambda^{2}\leq\frac{3\,\left(5\,\eta\,\lambda-4\right)^{2}}{2\,\left(7-8\,\eta\,\lambda\right)} \:\:\:\text{and}\:\:\:\:\: \frac{\tau_{c}}{\lambda}<0,
 \label{5.55}
\end{equation}whereas for
\begin{equation}
 \lambda^{2}<\frac{3\,\left(2-3\,\eta\,\lambda\right)}{2}\:\:\:\:\:\text{and/or}\:\:\:\:\: \frac{\tau_{c}}{\lambda}>0,
 \label{5.56}
\end{equation}it is a saddle point. On the other hand, for 
\begin{equation}
 \lambda^{2}>\frac{3\,\left(5\,\eta\,\lambda-4\right)^{2}}{2\,\left(7-8\,\eta\,\lambda\right)},
 \label{stablespiralconditionpointII.d}
\end{equation}$\mu_{1}$ and $\mu_{2}$ are complex with real part negative. If in addition we have $\tau_{c}/\lambda<0$ thus $\mu_{3}<0$. Since the determinant of the matrix of perturbations 
$\det{\mathcal{M}}|_{(x_{c},y_{c},u_{c})}=-\frac{27}{16}\,\frac{\tau_{c}}{\lambda}\,Y$ is negative, in this case the point II.c is a stable spiral (SS).
\begin{figure}[t]
\begin{center}
\begin{tabular}{lll}
\includegraphics[width=0.45\textwidth]{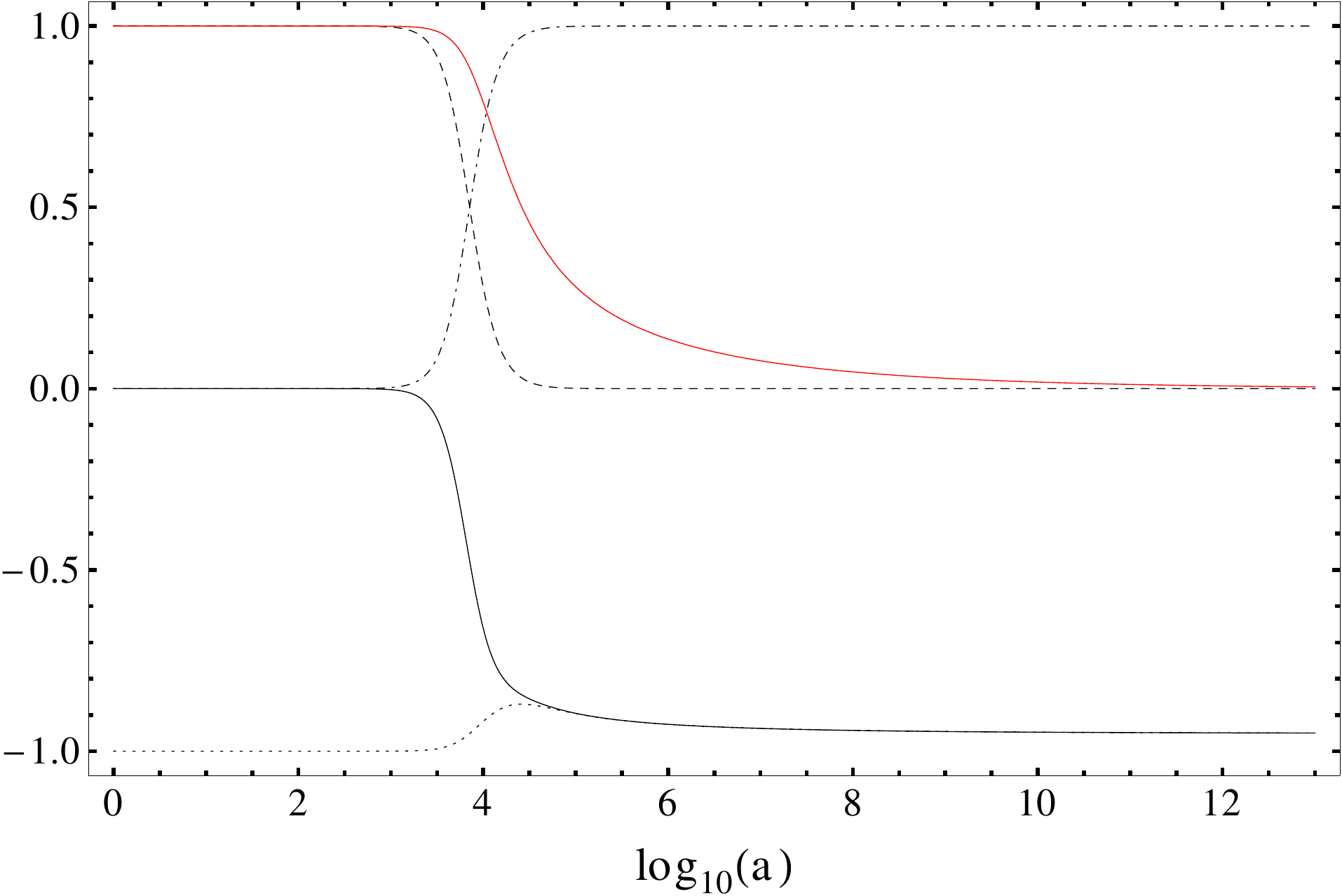}
& \includegraphics[width=0.45\textwidth]{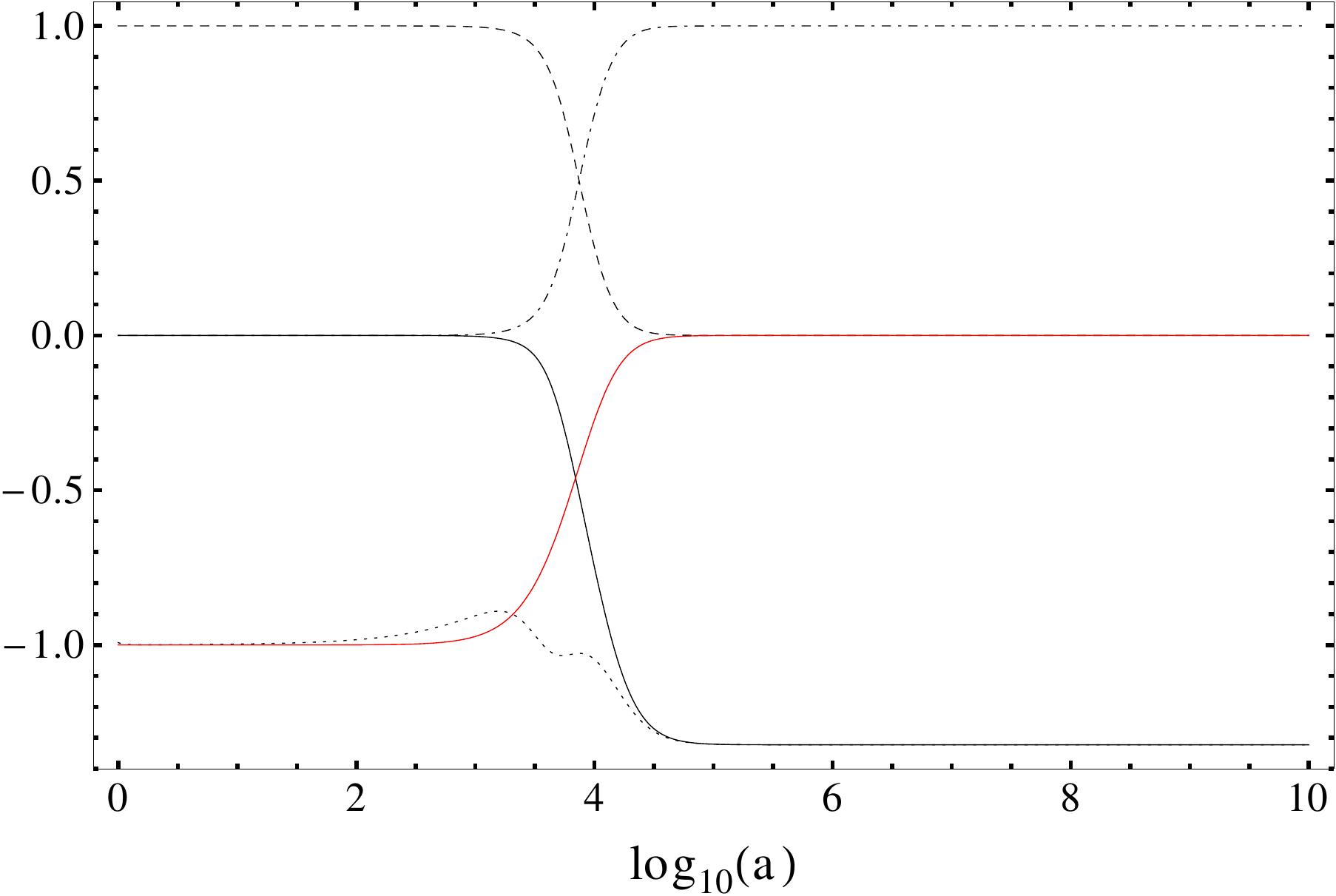}\\
\end{tabular}
\end{center}
\caption{Evolution of $\Omega_{m}$ (dashed), $\Omega_{\phi}$ (dotdashed), $\omega_{\phi}$ (dotted) and $\omega_{eff}$ (solid), with $\gamma=1$. In the left panel, for $\alpha(u)=1-u$ (red line, starting at $\alpha=1$),  we have chosen  $\lambda=0.8$ and $\eta=-0.2$.
The present epoch ($N=\log_{10}{(a)}\approx4$) corresponds to $\Omega_{\phi}\approx0.72$, $\Omega_{m}\approx0.28$, $\omega_{\phi}\approx-0.92$ and $\omega_{eff}\approx-0.66$. The Universe asymptotically evolves toward $\Omega_{\phi}=1$ and $\omega_{\phi}=\omega_{eff}=-0.95$.
In the right panel, for $\alpha(u)=-1+u$ (red line, starting at $\alpha=-1$), we have chosen $\lambda=0.3$ and $\eta=-1.4$.
At the present time, also with $\Omega_{\phi}\approx0.72$ and $\Omega_{m}\approx0.28$, we have  $\omega_{\phi}\approx-1.05$ and $\omega_{eff}\approx-0.75$. In this case, the evolution converges to $\Omega_{\phi}=1$ and $\omega_{\phi}=\omega_{eff}=-1.32$.}
\label{figure2}
\end{figure}
Finally, for $\eta\,\lambda<1-\frac{\lambda^{2}}{6}$, the point II.d is a stable node if
\begin{equation}
 \eta\,\lambda<\frac{2\,\left(3\,\gamma-\lambda^{2}\right)}{3\,\left(\gamma+2\right)}\:\:\:\:\:\:\: \text{and}\:\:\:\:\:\: \left(\lambda+3\,\eta\right)\,\tau_{c}<0,
 \label{5.57}
\end{equation}otherwise it is a saddle point. Whenever accelerated expansion occurs, $\eta\,\lambda<\frac{1}{2}-\frac{\lambda^{2}}{4}$, (and satisfying the constraint \eqref{5.57}) this fixed point is a stable node and therefore an attractor. 
Just like the point I.c, the fixed point II.d is also a viable cosmological solution to explain the current phase of accelerated expansion. 
In Fig. \ref{figure2} the Universe tends asymptotically to the solution II.d for dynamically changing $\alpha(u)$ with accelerated expansion, first passing through the matter-dominated solution I.a with constant $\alpha$.
In the left panel, we consider the function $f(\phi)=\frac{1}{\kappa}\,\left(1-e^{-\kappa\,\phi}\right)$  such that  $\alpha(u)=1-u$ and $\tau_{c}=-1$.  Similarly, in the right panel, it is considered the function $f(\phi)=\frac{1}{\kappa}\,\left(1+e^{\kappa\,\phi}\right)$ such that  $\alpha(u)=-1+u$ and $\tau_{c}=1$. 

\section{Concluding remarks}

A novel model has been proposed, named ``new teleparallel dark energy'' (NTDE), in which it is allowed a nonminimal coupling between the quintessence field and torsion. As is well-known \cite{JGPereira,12}, a scalar field couples to torsion through its four-derivative---which is a vector field. It is then natural to consider a nonminimal coupling of the four-derivative of the scalar field and the vector part of torsion or ``vector torsion''. 
It is important to note that unlike the TDE scenario of Ref.~\cite{10}, which was proposed by following an analogy with the nonminimally coupled scalar field in GR, in the present model the nonminimal coupling is proposed according to a conceptually different description of the gravitational field in TG \cite{JGPereira,12,SelectedtopicsinT}. In a generalization of TG which is consistent with local spacetime kinematics regulated by the de Sitter group SO(1,4), a cosmological function naturally emerges such that its four-derivative is nonminimally coupled with the vector torsion \cite{HJennenJGPereira}. Moreover, once the nonminimal coupling is switched on, the scalar field becomes coupled, through its four-derivative, to the spin tensor of the gravitational field, at the field equations level. However, the coupling to the gravitational spin tensor becomes negligible on cosmological scales (Eq. \eqref{C.16}).

By studying the dynamics of the model, we have found the critical points, presented in Tables \ref{table1} and \ref{table3}. In Table \ref{table1} the final attractor of the Universe is a dark-energy-dominated de Sitter solution I.c, with $\Omega_{\phi}=1$ and $\omega_{\phi}=\omega_{eff}=-1$. On the other hand, in Table \ref{table3}, the final attractor is a scalar-field-dominated solution II.d, also with $\Omega_{\phi}=1$, but in this case either $\omega_{\phi}=\omega_{eff}\geq-1$ or $\omega_{\phi}<-1$ in which case it represents a phantom-type solution. 
However, unlike the dark energy models with phantom (ghost) field \cite{1, DarkETO}, here the phantom regime ($\omega_{\phi}<-1$) is described without the problem of quantum instability \cite{QuantumInstability}. Additionally, unlike of the TDE scenario, here the phantom Universe is an attractor solution for the cosmological dynamics.
The fixed points I.c and II.d are viable cosmological solutions to explain the current accelerated expansion of the Universe.

It is interesting to remark that the models TDE and NTDE are mathematically related through a conformal transformation. This can be seen by defining transformed scalar field and potential, and by adding an explicit coupling between the scalar field and matter.
However, as already pointed out in the case of scalar-tensor theories and coupled dark energy in GR, where this type of mathematical relationship also occurs \cite{4,31,ConformalTransformation,JordanvsEinstein}, the two models are physically different. 
Furthermore, differently from coupled dark energy in GR, here we have taken the freedom to exclude a possible explicit coupling between the scalar field and matter (although this could be considered in a future work).

\section{Acknowledgments}

The author would like to thank J. G. Pereira and A. A. Deriglazov for useful discussions and suggestions. He would like to thank also CAPES(Program PNPD) for financial support.

\appendix\label{App1}

\section{Matrix of linear perturbations} 
\label{Components}

\subsection{Constant $\alpha$}
We find the follows components for the matrix of linear perturbations $\mathcal{M}$
\begin{multline}
\mathcal{M}_{11}=\frac{\eta\,u_{c}\,\left(\lambda\,y_{c}^{2}-\sqrt{6}\,\left(2\,\alpha\,\eta-3\,\gamma+6\right) \,x_{c}\right)-\gamma\,y_{c}^{2}+3\,\left(2\,\alpha\,\eta-\gamma+2\right) \,x_{c}^{2}-3\,\left(2-\gamma\right)}{{\eta}^{2}\,u_{c}^{2}+\frac{2}{3}}+\\
3\,\left(2-\gamma\right), 
\label{5.30}
\end{multline}
\begin{equation}
\mathcal{M}_{12}=\frac{\left(2\,\left(\eta\,\lambda\,u_{c}-\gamma\right)\,x_{c}+\sqrt{6}\,\gamma\,\eta\,u_{c}+\frac{2\,\sqrt{6}}{3}\,\lambda\right) \,y_{c}}{{\eta}^{2}\,u_{c}^{2}+\frac{2}{3}},
\label{5.31}
\end{equation}
\begin{multline}
\mathcal{M}_{13}=-\frac{\eta\,\left(\left(2\,\lambda\,x_{c}+\sqrt{6}\,\gamma\right)\,y_{c}^{2}-\sqrt{6}\,\left(2\,\alpha\,\eta-3\,\gamma+6\right) \,x_{c}^{2}+\sqrt{6}\,\left(2-\gamma\right)\right)}{2\,\left(\eta^{2}\,u_{c}^{2}+\frac{2}{3}\right)}+\\
\frac{2\,\eta}{3}\,[\eta\,u_{c}\,\left(\left(3\,\gamma\,x_{c}-\sqrt{6}\,\lambda\right) \,y_{c}^{2}-3\,\left(2\,\alpha\,\eta-\gamma+2\right)\,x_{c}^{3}+9\,\left(2-\gamma\right)\,x_{c}\right)+\\
\left(2\,\lambda\,x_{c}+\sqrt{6}\,\gamma\right)\,y_{c}^{2}-\sqrt{6}\,\left(2\,\alpha\,\eta-3\,\gamma+6\right)\,x_{c}^{2}+\sqrt{6}\,\left(2-\gamma\right)]\,\left({\eta}^{2}\,u_{c}^{2}+\frac{2}{3}\right)^{-2},
\label{5.32}
\end{multline}
\begin{equation}
\mathcal{M}_{21}=\frac{\left(2\,\left(2\,\alpha\,\eta-\gamma+2\right)\,x_{c}-\sqrt{6}\,\left(2-\gamma\right)\,\eta\,u_{c}\right)\,y_{c}}{\eta^{2}\,u_{c}^{2}+\frac{2}{3}}-\frac{\sqrt{6}\,\lambda\,y_{c}}{2},
\label{5.33}
\end{equation}
 \begin{equation}
\mathcal{M}_{22}=\frac{\eta\,u_{c}\,\left(3\,\lambda\,y_{c}^{2}-\sqrt{6}\,\left(2-\gamma\right)\,x_{c}\right)-3\,\gamma\,y_{c}^{2}+\left(2\,\alpha\,\eta-\gamma+2\right)\,x_{c}^{2}+\gamma-2}{\eta^{2}\,u_{c}^{2}+\frac{2}{3}}-\frac{\sqrt{6}\,\lambda\,x_{c}-6}{2},
\label{5.34}
\end{equation}
\begin{multline}
\mathcal{M}_{23}=\frac{2\,\eta\,y_{c}\,\left(3\,\eta\,u_{c}\,\left(\gamma\,y_{c}^{2}+\left(-2\,\alpha\,\eta+\gamma-2\right)\,x_{c}^{2}-\gamma+2\right)+2\,\lambda\,y_{c}^{2}-2\,\sqrt{6}\,\left(2-\gamma\right)\,x_{c}\right)}{3\,{\left(\eta^{2}\,u_{c}^{2}+\frac{2}{3}\right)}^{2}}-\\
\frac{\eta\,y_{c}\,\left(\lambda\,y_{c}^{2}-\sqrt{6}\,\left(2-\gamma\right)\,x_{c}\right)}{\eta^{2}\,u_{c}^{2}+\frac{2}{3}},
\label{5.35}
\end{multline}
 \begin{equation}
\mathcal{M}_{31}=\sqrt{6}\,\alpha, \:\:\:\:\:\: \mathcal{M}_{32}=0, \:\:\:\:\:\: \mathcal{M}_{33}=0.
\label{5.36}
\end{equation}

\subsection{Dynamically changing $\alpha(u)$}

For dynamically changing $\alpha(u)$ the components of the matrix of perturbation $\mathcal{M}$ are written as
\begin{equation}
\mathcal{M}_{11}=\frac{3\,\left(\eta\,\left( \lambda\,y_{c}^{2}-3\,\sqrt{6}\,\left(2-\gamma\right)\,x_{c}\right)-\gamma\,y_{c}^{2}+3\,\left(2-\gamma\right)\,\left( x_{c}^2-1\right)\right) }{3\,\eta^{2}+2}+3\,\left(2-\gamma\right),
\label{5.43}
\end{equation}
\begin{equation}
\mathcal{M}_{12}=\frac{\sqrt{6}\,\left(\sqrt{6}\,\left(\eta\,\lambda-\gamma\right)\,x_{c}+2\,\lambda+3\,\gamma\,\eta\right) \,y_{c}}{3\,\eta^{2}+2},
\label{5.44}
\end{equation}
\begin{multline}
\mathcal{M}_{13}=-\frac{1}{2}\,[3\,\eta\,\left(\left(2\,\lambda\,x_{c}+\sqrt{6}\,\gamma\right)\,y_{c}^{2}-4\,\tau_{c}\,x_{c}^{3}-\sqrt{6}\,\left(2-\gamma\right) \,\left(3\,x_{c}^{2}-1\right)\right)-\\
4\,\left(3\,\gamma\,x_{c}-\sqrt{6}\,\lambda\right)\,y_{c}^{2}-4\,\sqrt{6}\,\tau_{c}\,x_{c}^{2}+12\,\left(2-\gamma\right)\,x_{c}\,\left(x_{c}^{2}-3\right)]\,\left(3\,\eta^{2}+2\right)^{-1}+\\
2\,[3\,\eta\,\left(\left(2\,\lambda\,x_{c}+\sqrt{6}\,\gamma\right) \,y_{c}^{2}-\sqrt{6}\,\left(2-\gamma\right)\,\left(3\,x_{c}^{2}-1\right)\right)-2\,\left(3\,\gamma\,x_{c}-\sqrt{6}\,\lambda\right) \,y_{c}^{2}+\\
6\,\left(2-\gamma\right)\,x_{c}\,\left(x_{c}^{2}-3\right)]\,\left(3\,\eta^{2}+2\right)^{-2}-\sqrt{6}\,\tau_{c}\,x_{c}^{2},
\label{5.45}
\end{multline}
\begin{equation}
\mathcal{M}_{21}=\frac{3\,\left(2-\gamma\right)\,\left(2\,x_{c}-\sqrt{6}\,\eta\right)\,y_{c}}{3\,\eta^{2}+2}-\frac{\sqrt{6}\,\lambda\,y_{c}}{2},
\label{5.46}
\end{equation}
\begin{equation}
\mathcal{M}_{22}=\frac{3\,\left(\eta\,\left(3\,\lambda\,y_{c}^{2}-\sqrt{6}\,\left(2-\gamma\right)\,x_{c}\right)-3\,\gamma\,y_{c}^{2}+\left(2-\gamma\right)\,\left(x_{c}^{2}-1\right)\right)}{3\,\eta^{2}+2}-\frac{\sqrt{6}\,\lambda\,x_{c}-6}{2},
\label{5.47}
\end{equation}
\begin{multline}
\mathcal{M}_{23}=\frac{12\,y_{c}\,\left(\eta\,\left(\lambda\,y_{c}^{2}-\sqrt{6}\,\left(2-\gamma\right)\,x_{c}\right)-\gamma\,y_{c}^{2}+\left(2-\gamma\right)\,\left(x_{c}^{2}-1\right)\right)}{{\left(3\,\eta^{2}+2\right) }^{2}}+\\
\frac{3\,y_{c}\,\left(\eta\,\left(-\lambda\,y_{c}^{2}+2\,\tau_{c}\,x_{c}^{2}+\sqrt{6}\,\left(2-\gamma\right)\,x_{c}\right)+2\,\gamma\,y_{c}^{2}-2\,\left(2-\gamma\right)\,\left(x_{c}^{2}-1\right)\right)}{3\,\eta^{2}+2},
\label{5.48}
\end{multline}
\begin{equation}
\mathcal{M}_{31}=\mathcal{M}_{32}=0, \:\:\:\:\:\:\:\: \mathcal{M}_{33}=\sqrt{6}\,\tau_{c}\,x_{c}.
\label{5.49}
\end{equation}
Here $\tau_{c}$ is defined by $\tau_{c}\equiv\frac{d\alpha(u)}{du}|_{u=u_{c}}$.

\end{document}